\documentclass[noshowpacs,amsmath,
twocolumn,
superscriptaddress,
8pt,
aps,prb]{revtex4-1}
\usepackage{setspace}
\usepackage{amsmath}
\usepackage{graphicx}
\usepackage[nearskip,margin = 0pt]{subfig}

\usepackage{verbatim}
\usepackage{amsfonts}
\usepackage{amssymb}
\usepackage{epstopdf} 
\usepackage{xcolor}
\usepackage{verbatim}
\usepackage{multibib}
\DeclareGraphicsExtensions{.pdf,.eps,.png,.jpg,.mps} 
\begin{document}
\title{Imaging soliton dynamics in optical microcavities}

%\title{Real-time characterization of Kerr solitons dynamics using comb-based temporal interferometry}

\author{Xu Yi$^{\ast}$, Qi-Fan Yang$^{\ast}$, Ki Youl Yang, and Kerry Vahala$^{\dagger}$\\
T. J. Watson Laboratory of Applied Physics, California Institute of Technology, Pasadena, California 91125, USA.\\
$^{\ast}$These authors contributed equally to this work.\\
$^{\dagger}$Corresponding author: vahala@caltech.edu}

\date{\today}

\maketitle

%%%%%%%%%%%%%Start of maintext%%%%%%%%%
\noindent {\bf Solitons are self-sustained wavepackets that occur in many physical systems. Their recent demonstration in optical microresonators has provided a new platform for study of nonlinear optical physics with practical implications for miniaturization of time standards, spectroscopy tools and frequency metrology systems. However, despite its importance to  understanding of soliton physics as well as development of new applications, imaging the rich dynamical behaviour of solitons in microcavities has not been possible.  These phenomena require a difficult combination of high-temporal-resolution and long-record-length in order to capture the evolving trajectories of closely-spaced microcavity solitons. Here, an imaging method is demonstrated that visualizes soliton motion with sub-picosecond resolution over arbitrary time spans. A wide range of complex soliton transient behaviors are characterized in the temporal or spectral domain, including soliton formation, collisions, spectral breathing and soliton decay. This method can serve as a universal visualization tool for understanding complex soliton physics in microcavities.
}

\medskip
Temporal solitons are indispensable in optical fiber systems \cite{kivshar2003optical} and exhibit remarkable nonlinear phenomena \cite{dudley2014instabilities}.  The potential application of solitons to buffers and memories\cite{Wabnitz:93,leo2010temporal} as well as interest in soliton physics has stimulated approaches for experimental visualization of multi-soliton trajectories.  Along these lines, the display of solitons trajectories in a co-moving frame \cite{jang2013ultraweak} allows an observer to move with the solitons and is being used to monitor soliton control and interactions of all types in fiber systems\cite{jang2013ultraweak,luo2015real,jang2015temporal,jang2016controlled,herink2017real}. However, this useful data visualization method relies upon soliton pulse measurements that are either limited in bandwidth (pulse resolution) or record length.  It is therefore challenging to temporally resolve solitons over the periods often required to observe their complete evolution. For example, the time-lens method\cite{foster2008silicon} can provide the required femtosecond-resolution, but has a limited record length set by the pump pulse. Also, while the relative position of closely-spaced soliton complexes\cite{herink2017real} can be inferred over time from their composite DFT spectra\cite{Goda:2012}, Fourier inversion requires the constituent solitons to have similar waveforms which restricts the generality of the technique.

These limitations are placed in sharp focus by recent demonstrations of soliton generation in microcavities\cite{herr2014temporal,yi2015soliton,brasch2016photonic,wang2016intracavity,joshi2016thermally,obrzud2016temporal,lobanov2016harmonization}. This new type of dissipative soliton\cite{akhmediev2008dissipative} was long considered a theoretical possibility\cite{Wabnitz:93} and was first observed in optical fiber resonators \cite{leo2010temporal}. Their microcavity embodiment poses severe challenges for imaging of dynamical phenomena by conventional methods, because multi-soliton states feature inherently closely spaced solitons. Nonetheless, the compactness of these systems has tremendous practical importance for miniaturization of frequency comb technology\cite{diddams2010evolving} through chip-based microcombs\cite{del2007optical,kippenberg2011microresonator}. Indeed, spectroscopy systems\cite{suh2016microresonator}, coherent communication\cite{marin2017microresonator}, ranging\cite{trocha2017ultrafast,suh2017microresonator}, and frequency synthesis\cite{spencer2017integrated} demonstrations using the new miniature platform have already been reported. Moreover, the unique physics of the new soliton microcavity system has lead to observation of many unforeseen physical phenomena involving compound soliton states, such as Stokes solitons\cite{yang2017stokes}, soliton number switching\cite{wang2017universal} and soliton crystals\cite{cole2017soliton}. 

\begin{figure*}[!ht]
\captionsetup{singlelinecheck=off, justification = RaggedRight}
\includegraphics[width=18cm]{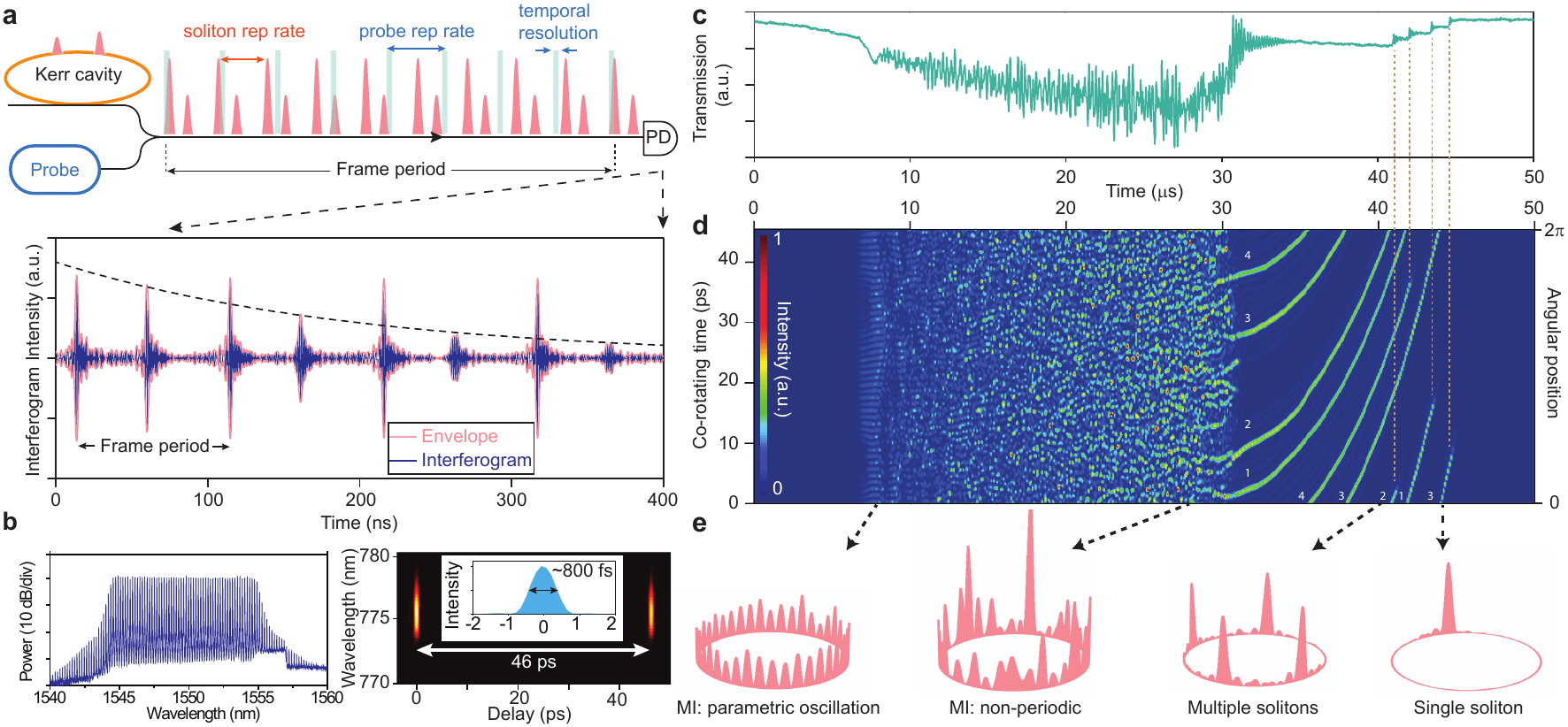}
\caption{{\bf Coherent sampling of dissipative Kerr soliton dynamics.} {\bf a,} Conceptual schematic showing microcavity signal (red) combined with the probe sampling pulse train (blue) using a bidirectional coupler. The probe pulse train repetition rate is offset slightly from the microcavity signal. It temporally samples the signal upon photo detection to produce an interferogram signal shown in the lower panel. The measured interferogram shows several frame periods during which two solitons appear with one of the solitons experiencing decay. {\bf b,} Left panel is the optical spectrum and right panel is the FROG trace of the probe EO comb (pulse repetition period is shown as 46 ps). An intensity autocorrelation in the inset shows a full-width-half-maximum pulse width of 800 fs. {\bf c,} Microresonator pump power transmission when the pump laser frequency scans from higher to lower frequency. Multiple ``steps" indicate the formation of solitons. {\bf d,} Imaging of soliton formation corresponding to the scan in panel {\bf c}. The x-axis is time and the y-axis is time in a frame that rotates with the solitons (full scale is one round-trip time). The right vertical axis is scaled in radians around the microcavity. Four soliton trajectories are labeled and fold-back into the cavity coordinate system. The color bar gives their signal intensity.  {\bf e,} Soliton intensity patterns measured at four moments in time are projected onto the microcavity coordinate frame. The patterns correspond to initial parametric oscillation\cite{kippenberg2004kerr} in the modulation instability (MI) regime\cite{Wabnitz:93,leo2010temporal}, non-periodic behavior (MI regime), four soliton and single soliton states\cite{leo2010temporal,herr2014temporal}.}
\label{figure1}
\end{figure*}

In this work, we report imaging of a wide range of soliton phenomena in microcavities. Soliton formation, collisions\cite{jang2016controlled}, breathing\cite{akhmediev1986modulation,leo2013dynamics,luo2015real,bao2016observation}, Raman shifting\cite{karpov2016raman,wang2017stimulated} as well as soliton decay are observed. Significantly, femtosecond-time-scale resolution over arbitrary time spans (distances) is demonstrated (and required) in these measurements.
Also, real-time spectrograms are produced along-side high-resolution soliton trajectories. These features are derived by adapting coherent linear optical sampling\cite{Coddington:09,ferdous2009dual,duran2015ultrafast} to the problem of microcavity solition imaging. To image the soliton trajectories, a separate optical probe pulse stream is generated at a pulse rate that is close to the rate of the solitons to be imaged in the microcavity. The small difference in these rates causes a pulse-to-pulse temporal shift of the probe pulses relative to the microcavity signal pulses as illustrated in fig. 1a. By heterodyne detection of the combined streams, the probe pulses coherently sample the microcavity signal producing a temporal interferogram\cite{coddington2009rapid,coddington2016dual} shown in fig. \ref{figure1}a. Ultimately, the time shift per pulse accumulates so that the sampling repeats in the interferogram at the ``frame rate" which is described below, and is close in value to the difference of sampling and signal rates. Probe pulses have a sub-picosecond temporal resolution that enables precise monitoring of the temporal location of the soliton pulses. Moreover, the coherent mixing of probe and soliton pulses allows extraction of each soliton's spectral evolution by fast Fourier transform of the interferogram. In principle, the probe pulses can be generated by a second microcavity soliton source operating in steady state. However, in the present measurement, an electro-optical (EO) comb is used\cite{murata2000optical,ferdous2009dual,duran2015ultrafast}. The EO comb pulse rate is conveniently adjusted electronically to match the rates of various phenomena being probed within the microcavity. %As now described, the extremely high repetition rates of solitons in microcavities translates to a high sampling rate that is used for high-resolution imaging of multi-soliton motion over kilometer-scale propagation distances. %These dissipative solitons\cite{akhmediev2008dissipative} were long considered a theoretical possibility\cite{Wabnitz:93} and have also been observed in optical fiber resonators \cite{leo2010temporal}. In addition to providing a new platform for investigation of nonlinear optical physics\cite{leo2013dynamics,jang2014observation,luo2015real,jang2015temporal,bao2016observation,brasch2016photonic,karpov2016raman,yang2017stokes,wang2017stimulated},
%Their realization in microcavities represents a major turning point for the subject of frequency microcombs\cite{del2007optical,kippenberg2011microresonator} where the resulting highly stable soliton mode locking has been applied to a range of microcomb systems\cite{suh2016microresonator,marin2017microresonator,trocha2017ultrafast,suh2017microresonator,spencer2017integrated}. 

%In addition,  it provides a new platform for investigation of nonlinear optical physics\cite{leo2013dynamics,jang2014observation,luo2015real,jang2015temporal,bao2016observation,brasch2016photonic,karpov2016raman,yang2017stokes,wang2017stimulated}.

%Here, the extremely high repetition rates of solitons in microcavities translates to a high sampling rate, which is used for high-resolution imaging of soliton motion over kilometer-scale distances. %Collisions\cite{jang2016controlled}, breathing\cite{akhmediev1986modulation,leo2013dynamics,luo2015real,bao2016observation}, Raman shifting\cite{karpov2016raman,wang2017stimulated} as well as other phenomena can accordingly be studied at an entirely new level of detail. 
%which is closely related to dual-comb spectroscopy \cite{coddington2016dual} and dual-comb ranging \cite{coddington2009rapid}. } Significantly, however, these advantages would not translate well to the problem of soliton imaging if not for the

%The pulse width of the EO comb determines the temporal resolution and the repetition rate of the comb pulses is slightly offset from the repetition rate of the microcavity signal to be measured.               

\begin{figure*}[!ht]
\captionsetup{singlelinecheck=off, justification = RaggedRight}
\includegraphics[width=18cm]{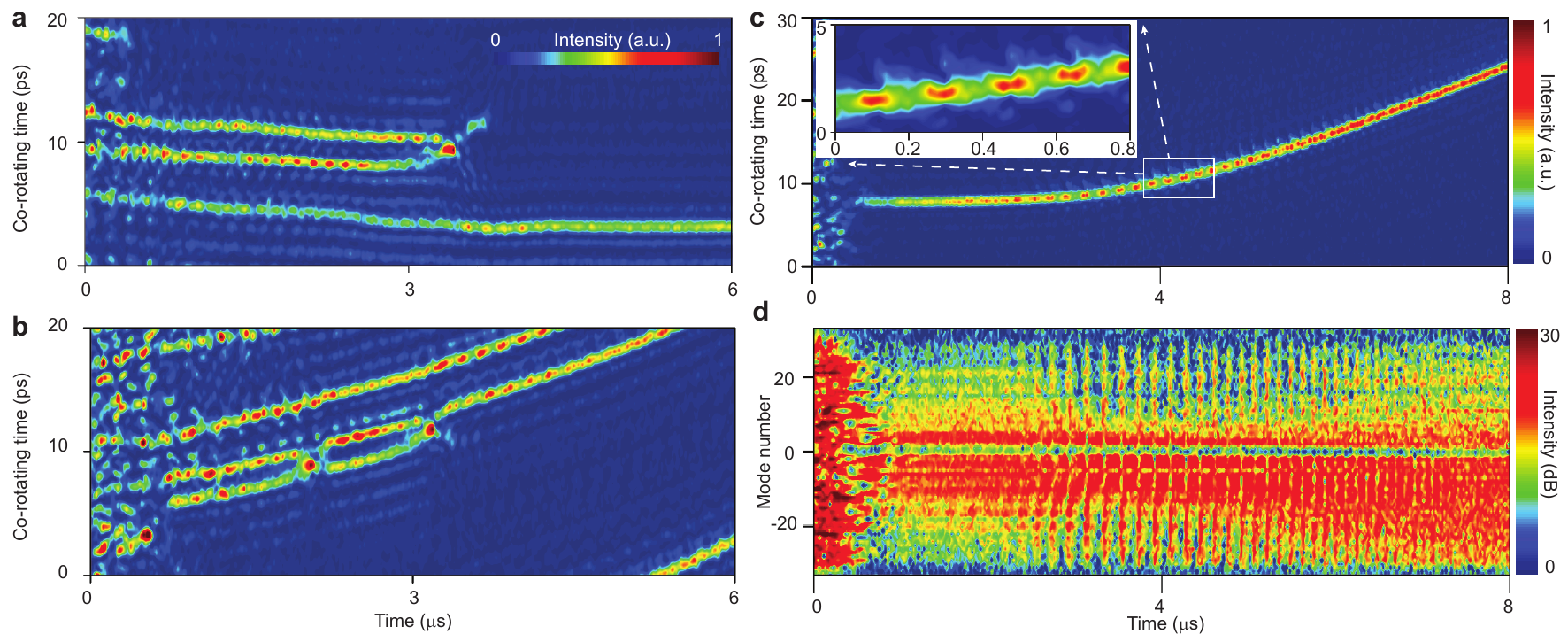}
\caption{{\bf Temporal and spectral measurements of non-repetitive soliton events.} {\bf a,} Two solitons collide and annihilate. {\bf b,} Two solitons survive a collision, but collide again and one soliton is annihilated. {\bf c,} Motion of a single soliton state showing peak power breathing along its trajectory. A zoom-in view of the white rectangular region is shown as the inset. {\bf d,} Spectral dynamics corresponding to panel {\bf c}. The y-axis is the relative longitudinal mode number corresponding to specific spectral lines of the soliton. Mode zero is the pumped microcavity mode. The soliton spectral width breaths as the soliton peak power modulates. The spectrum is widest when peak power is maximum. The frame rate is 50 MHz for all panels.}
\label{figure2}
\end{figure*}

The soliton signal is produced by a 3 mm diameter silica wedge resonator with FSR of 22 GHz and intrinsic quality factor above 200 million \cite{lee2012chemically,yi2015soliton}.  The device generates femtosecond soliton pulses when pumped at frequencies slightly lower than a cavity resonant frequency\cite{yi2015soliton}. To sample the 22 GHz soliton signal the EO comb was formed by modulation of a tunable continuous wave (CW) laser. The EO comb features $\sim$ 1.3 THz optical bandwidth (within 1 dB power variation) and an 800 fs FWHM pulse width measured by frequency-resolved optical gating and autocorrelation as shown in fig. \ref{figure1}b. A detailed schematic of the complete experimental setup is provided in the supplementary fig. S1. In all presented measurements, the pump laser of the resonator scans linearly from higher to lower frequency to initiate parametric oscillation in the microcavity followed by chaotic dynamics. Ultimately, ``step-like" features are observable in the resonator transmitted power (fig. \ref{figure1}c) indicating the formation of soliton states \cite{herr2014temporal}. The typical pump power and laser scan speed are $\sim$ 70 mW and $\sim -1$ MHz/$\mu$s, respectively.

As described above, heterodyne-detection of the soliton signal and the EO-comb pulse produces the electrical interferogram. The period of the signals in the interogram is adjusted by tuning the EO-comb repetition rate. In the initial measurements, it is set to $\sim$ 10 MHz lower than the rate of the microcavity signal so that the nominal period in the interferogram is $\sim$ 100 ns. %Because all microcavity signals of interest have rates that are close to the FSR of the microcavity, in setting the EO-comb repetition rate it is convenient to first precisely determine the microcavity FSR by modulation-side-band spectroscopy \cite{li2012sideband}. For these measurements a common laser is used to pump the EO-comb and the soliton source, however, the pumping frequency for the soliton source is offset using an acousto-optic modulator. 
To display the interferogram signal a co-rotating frame is applied. First, a frame period $T$ is chosen that is close to the period of signals of interest in the interferogram. Integer steps (i.e., $mT$) are plotted along the x-axis while the interferogram is plotted along the y-axis, but offset in time by the x-axis time step (i.e., $t - mT$). To make connection to the physical time scale of the solitons, the y-axis time scale is also compressed by the same bandwidth compression factor ($T \times$ FSR) that accompanies the sampling process. The y-axis scale is accordingly set to span one microcavity round-trip time.  A typical measurement plotted in this manner is given in fig. \ref{figure1}d. Because this way of plotting the data creates a co-rotating reference frame, a hypothetical soliton pulse with an interferogram period equal to the frame rate $T$ would appear as a horizontal line in fig. \ref{figure1}d. On the other hand, slower (higher) rate solitons would appear as lines tilted upward (downward) in the plot. In creating the imaging plot, a Hilbert transformation is applied to the interferogram followed by taking the square of its amplitude to produce a pulse envelope intensity profile. The vertical co-rotating time axis can be readily mapped into an image of the soliton angular position within the circular microcavity as shown in fig. \ref{figure1}d.

%Movies of the corresponding multi-soliton motion around the microcavity are also provided in the Supplementary Section.
%In reference to the earlier discussion, this arrangement of rates produces a frame rate and a bandwidth compression of 22,000 MHz / 10 MHz = 2200.
%The envelope of the single soliton state has a temporal full-width-half-maximum of $\sim 0.81$ ps, which is limited by the EO sampling pulse width. 

Imaging of soliton formation and multi-soliton trajectories is observable in fig. \ref{figure1}d. For comparison with the transmitted power, the time-axis scale is identical in fig. \ref{figure1}c and fig. \ref{figure1}d. As the pump laser frequency initially scans towards the microcavity resonant frequency its coupled power increases. At $\sim 8$ $\mu$s the resonator enters the modulation instability regime \cite{Wabnitz:93,leo2010temporal,herr2014temporal}. Initially, a periodic temporal pattern is observable in fig. \ref{figure1}d corresponding to parametric oscillation\cite{kippenberg2004kerr}.
%The frame interval $T$ was adjusted slightly so that these oscillations appear as horizontal lines in the figure. With further tuning of pumping and increase of coupled power to the resonator, 
Soon after, the cavity enters a regime of non-periodic oscillation. At $\sim 31$ $\mu$s, this regime suddenly transitions into four soliton pulses. The soliton positions evolve with scan time and disappear one-by-one. 
%Relative motion of the solitons occurs at the beginning of soliton formation. In addition, 
All solitons have upward curved trajectories, showing that the soliton repetition rate decreases as the scan progresses. This soliton rate shift is caused by the combination of the Raman self-frequency shift effect and anomalous dispersion in the silica resonator \cite{karpov2016raman,yi2016theory} and a similar effect on soliton trajectory is observed in optical fiber cavities\cite{wang2017stimulated}. 
The features of soliton formation and evolution observed in fig. \ref{figure1}d compare well with numerical simulations presented in Supplementary fig. S2. Moreover, relative soliton positions can be extracted from the interferogram measurement (Supplementary fig. S3) and illustrate solitons stabilizing their relative positions. Movies of the corresponding multi-soliton motion around the microcavity are also provided in the Supplementary Section. Finally, the cavity states at four moments in time are plotted within the circular microcavity in fig. \ref{figure1}e. These correspond to parametric oscillation, non-periodic modulational instability, four soliton and single soliton states.

%The vertical co-rotating time axis can be readily mapped into an image of the soliton angular position within the circular microcavity as shown in fig. \ref{figure1}e. In creating these plots, a Hilbert transformation is applied to the interferogram followed by taking the square of its amplitude to produce a pulse envelope intensity profile. Movies of the corresponding multi-soliton motion around the microcavity are also provided in the Supplementary Section.

\begin{figure}[!ht]
\captionsetup{singlelinecheck=off, justification = RaggedRight}
\includegraphics{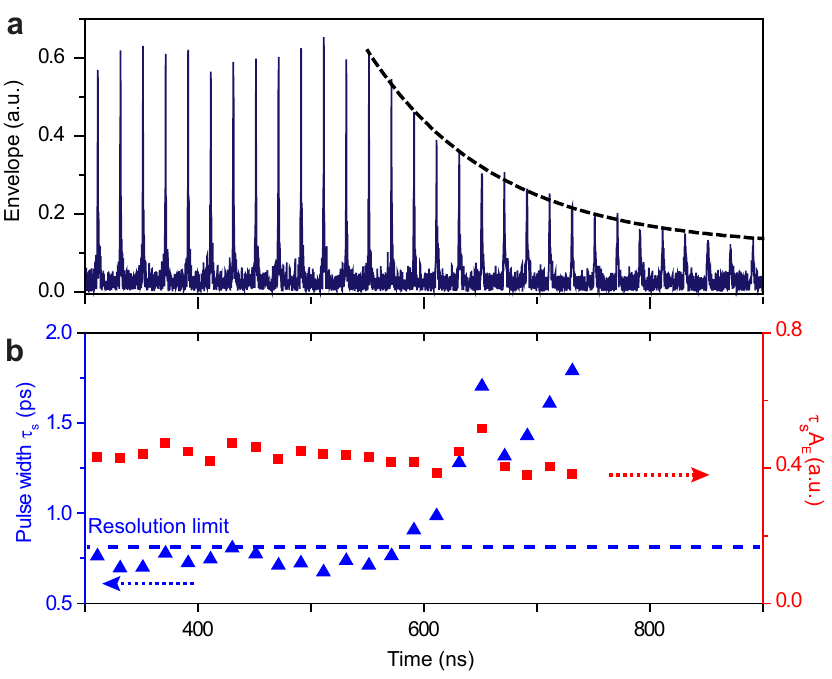}
\caption{{\bf Characterization of soliton decay.} {\bf a,} Interferogram envelope showing a single soliton experiencing decay. An exponential fitting is given as the dashed black line. {\bf b,} The measured pulse width (blue) is plotted versus time and its resolution limit is set by the EO comb pulse width. The product of soliton amplitude and pulse width is plotted in red.}
\label{figure4}
\end{figure}

A variety of non-repetitive multi and single soliton phenomena were measured in both temporal and spectral domains. To enable more rapid imaging the repetation rate of EO comb was adjusted to produce an interferogram at a rate of approximately 50 MHz. The frame period, $T$, was then reduced accordingly to approximately 20 nsec. Fig. \ref{figure2}a-b present observations of two solitons interacting. Soliton annihilation is observed in fig. \ref{figure2}a, wherein two solitons move towards each other, collide, create an intense peak upon collision and then disappear. A new phenomena, a ``wave splash", is observed immediately following the collision. In fig. \ref{figure2}b, two solitons collide but quickly recover and then collide again, after which point one soliton is annihilated. Significantly, the observation of these complex motions requires measurement of events in close temporal proximity over long time spans. Figure \ref{figure2}c-d shows measurement of a breathing soliton \cite{akhmediev1986modulation} in both the temporal and frequency domains. The spectrogram is obtained by applying a Fourier transform to the interferogram signal\cite{coddington2016dual}. The spectrum is widest when the soliton has maximum peak power. As an observation unrelated to the breathing action, the soliton spectral envelope in fig. \ref{figure2}d is continuously red shifted in frequency by the Raman self-frequency shift \cite{karpov2016raman,yi2015soliton} as its average power increases (increasing time in the plot).

Finally, soliton decay is analyzed using the sampling method. The measurement results are shown in fig. \ref{figure4}. In the experiment, the pump laser frequency is continuously tuned towards lower frequencies. After soliton formation, at some point the cavity-laser frequency detuning exceeds the soliton existence range and the soliton decays \cite{herr2014temporal,yi2015soliton}. Fig. \ref{figure4}a shows the interferogram signal just before and during the decay. Pulse widths ($\tau_s$) are extracted during the decay process and are plotted in fig. \ref{figure4}b. Also plotted in fig. \ref{figure4}b is the product of pulse width and soliton peak amplitude ($A_E$). Curiously, the soliton pulse width and peak amplitude preserve the same soliton product relationship as prior to decay. This is an indication that the decaying soliton pulse in the microcavity is constantly adapting itself to maintain the soliton waveform. A similar behavior is known to occur for conventional solitons in optical fiber \cite{agrawal2007nonlinear}. To the authors knowledge, this is the first time this behavior has been observed in real time. In the Method section the amplitude decay of the soliton in the interferogram trace is analyzed to extract a decay time and the cavity Q factor.

Coherent sampling induces a large bandwidth compression of the ultrafast signal that is equal to the sampling rate divided by the difference in the signal rate and the sampling rate. This compression is well known in the related techniques of dual comb spectroscopy \cite{coddington2016dual} and dual comb ranging \cite{coddington2009rapid}, and is also present in sampling of optical signals by four-wave mixing in optical fibers \cite{andrekson1991picosecond}. 
%,nelson1991optical,westlund2005high
In order to avoid spectral folding, the compressed signal bandwidth must lie within half of the EO comb sampling rate \cite{coddington2009rapid,coddington2016dual} (the Nyquist condition for sampling). As shown in the Method section, this basic condition establishes the following relationship between temporal resolution ($\tau$), frame rate ($f$) and the sampling rate (approximately the microcavity free-spectral-range, $\rm{FSR}$): $f < \tau \rm{FSR}^2/2$. This condition also reveals the quadratic importance of high sampling rates (equivalently large FSRs and correspondingly large soliton repetition rates) to create fast frame rates. In the current system, a temporal resolution of less than 1 ps combined with a 22 GHz sampling rate can enable frame rates as high as 200 MHz.

%\vspace{6pt}
%\noindent{\bf Summary.} 
Imaging of nonlinear dynamical phenomena including complex soliton interactions with high temporal/spatial resolution over arbitray time/length spans has been demonstrated. The temporal resolution in the current experiment is limited to $800$ fs, however, resolution at the 10s of fs level is possible by spectrally broadening the EO comb \cite{carlson2017ultrafast} used for coherent sampling. It is also possible to replace the EO-comb with a microcomb that is closely matched to the FSR of a microcavity to be sampled. Such matching has been recently used to implement dual soliton microcomb spectroscopy measurements\cite{suh2016microresonator}. In this case, even higher sampling rates would be possible that would enable GHz-scale frame rates. The coherent sampling method can serve as a general real-time state visualization tool to monitor the dynamics of microcavity systems. It would provide an ideal way to monitor the formation and evolution of soliton complexes such as Stokes solitons\cite{yang2017stokes}, soliton number switching\cite{wang2017universal} and soliton crystals\cite{cole2017soliton}. It can also be used to monitor the state of chip-based optical memories based on microresonator solitons. 
%It can also serve as a powerful tool for development of compact optical buffers and memories using microcavities.  

\medskip

\noindent\textbf{Methods}

\medskip

\begin{footnotesize}
\noindent{\bf Time constant in soliton decay.} 
In the soliton decay process, the average intracavity energy decays exponentially and its time constant equals the dissipation rate of the cavity ($\kappa = \omega/Q$), where $\omega$ is the optical frequency and Q is the loaded cavity Q factor. For large cavity-laser frequency detuning \cite{herr2014temporal,yi2016theory}, the average intracavity energy is approximately the soliton energy, $\tau_s A_E^2$, such that 
\begin{equation}
\tau_s(t) A_E^2(t) = \tau_s(0) A_E^2(0) e^{-\kappa t}.
\end{equation}
When the dissipation rate is relatively small compared to soliton Kerr nonlinear shift, the dissipation is a perturbation and the pulse maintains its soliton waveform \cite{agrawal2007nonlinear}. The corresponding balance of dispersion and Kerr nonlinearity requires that the product of soliton amplitude and pulse width be constant. This condition was also verified experimentally in figure 3b\cite{agrawal2007nonlinear,yi2016theory}, 
\begin{equation}
\tau_s(t) A_E(t)=\tau_s(0) A_E(0).
\end{equation}
Inserting eq. (2) into eq. (1) gives, 
\begin{equation}
A_E(t)=A_E(0)e^{-\kappa t}, \tau_s(t)=\tau_s(0)e^{\kappa t}, A_E^2(t)=A_E^2(0)e^{-2\kappa t}.
\end{equation}
In particular, the soliton amplitude decays at the cavity dissipation rate, the pulse width exponentially grows, and the soliton peak power decays twice as fast as the cavity dissipation rate. In the experiment, the fitted decay constant of the soliton amplitude is 133 ns, which corresponds to $\kappa/(2\pi) = 1.2$ MHz giving $Q= 161$ million. This value is in reasonable agreement with the measured loaded-Q factor of 140 million. 

\medskip

\noindent{\bf Nyquist condition for sampling.} 
In the EO comb sampling process the optical to electrical conversion is accompanied by a large bandwidth compression of the sampled signal. In effect, sampling stretches the time scale so that, for example, the optical temporal resolution ($\tau$) is stretched to $\tau \times {\rm FSR} / f$ after conversion to the electrical signal where $f$ is the frame rate given by $f \approx {\rm FSR} - f_{comb}$. This stretching means that the THz EO comb resolution bandwidth is compressed to an electrical bandwidth of $f/(\tau {\rm FSR})$. To avoid nonsensical signals in the electrical spectrum, the compressed bandwidth should lie within the Nyquist frequency set by the FSR \cite{coddington2009rapid}. This gives the condition $f/(\tau FSR) < {\rm FSR}/2$, or $f < \tau {\rm FSR}^2/2$. In practice, when the oscilloscope bandwidth ($f_\mathrm{osc}$) is smaller than the Nyquist frequency, the interferogram signal will be limited by the oscilloscope instead of the Nyquist frequncy, such that $f/(\tau {\rm FSR}) < f_\mathrm{osc}$, or $f < \tau f_\mathrm{osc} {\rm FSR}$. This is, in fact, the case in the present measurement as the oscilloscope bandwidth is 4 GHz while the Nyquist frequency is 11 GHz. In addition, the frequency components of the interferogram signal must be positive to avoid frequency folding near zero frequency. This requires that the carrier frequency of the interferogram signal is larger than half of the electrical bandwidth. In the present measurement, the carrier frequency is the frequency offset between the EO comb pump laser and the microcavity pump laser (defined as $\Delta \Omega$). As a result, this condition is expressed as $\Delta \Omega > f/(2\tau {\rm FSR})$.

\end{footnotesize}

\medskip

{\noindent \bf Data availability.} The data that support the plots within this paper and other findings of this study are available from the corresponding author upon reasonable request.

\medskip

\noindent\textbf{Acknowledgement}

\noindent The authors thank Stephane Coen and Yun-Feng Xiao for helpful comments during the preparation of this manuscript and gratefully acknowledge the Air Force Office of Scientific Research (AFOSR), NASA and the Kavli Nanoscience Institute.

\bibliography{main}  

%%%%%%%%%% Merge with supplemental materials %%%%%%%%%%
\pagebreak
\widetext
\begin{center}
\textbf{\large Supplemental Information: Imaging soliton dynamics in optical microcavities}
\end{center}
%%%%%%%%%% Merge with supplemental materials %%%%%%%%%%
%%%%%%%%%% Prefix a "S" to all equations, figures, tables and reset the counter %%%%%%%%%%
\setcounter{equation}{0}
\setcounter{figure}{0}
\setcounter{table}{0}
\makeatletter
\renewcommand{\theequation}{S\arabic{equation}}
\renewcommand{\thefigure}{S\arabic{figure}}
\renewcommand{\bibnumfmt}[1]{[S#1]}
\renewcommand{\citenumfont}[1]{S#1}
\begin{center}
Xu Yi$^{\ast}$, Qi-Fan Yang$^{\ast}$, Ki Youl Yang, and Kerry Vahala$^{\dagger}$\\
T. J. Watson Laboratory of Applied Physics, California Institute of Technology, Pasadena, California 91125, USA.\\
$^{\ast}$These authors contributed equally to this work.\\
$^{\dagger}$Corresponding author: vahala@caltech.edu
\end{center}

\section*{Supplementary Note 1: Experimental setup}

\begin{figure}[!ht]
\includegraphics[width=16.0cm]{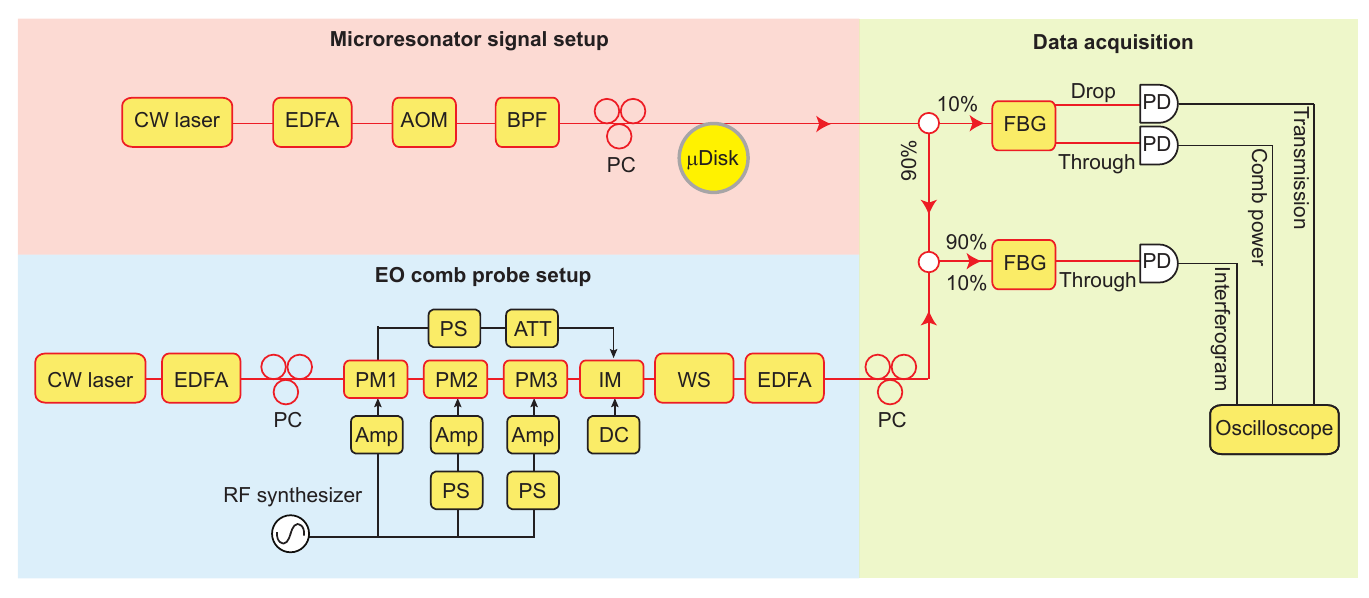}
\caption{{\bf Experimental setup.} CW laser: continuous-wave laser; EDFA: erbium-doped-fiber-amplifier; AOM: acousto-optic modulator; BPF: bandpass filter; PC: polarization controller; PM: phase modulator; IM: intensity modulator; PS: phase shifter; ATT: attenuator; Amp: RF amplifier; DC: DC voltage source; WS: optical waveshaper; FBG: fiber-Bragg-gating; PD: photodetector. 
%Note that the interferogram signal is detected using a 50 GHz fast photodetector.
}
\label{setup}
\end{figure}

Fig. S1 divides the experimental setup into three sections. In the microresonator section, a tunable, continuous-wave (cw) laser is used to pump the microcavity for production of solitons. An erbium-doped fiber amplifier (EDFA) amplifies its power to 500 mW and an acousto-optic modulator (AOM) is used for rapid control of power to the microcavity. A tunable bandpass filter (BPF) is used to block the spontaneous emission noise from the EDFA. The pump is coupled into the microcavity through a tapered-fiber \cite{slee2012chemically}. The emitted power from the microcavity (along with transmitted pump power) is split by a 90/10 fiber coupler. 10 percent of the power is sent to a fiber-Bragg grating (FBG) filter to separate the pump power and the microcomb power. The drop port output is the pump power transmission, while the through-port output is the comb power. Both the pump transmission and the microcomb power are detected with photodetectors (125 MHz bandwidth). The other 90 percent of the power is combined with the electro-optic (EO) modulation comb sampling pulse using a second fiber coupler.

In the EO comb setup, a pump laser is amplified by an EDFA to 200 mW and then phase modulated by three tandem lithium niobate modulators. The EO comb and microcavity setup can share the same pump laser when the acousto-optic modulator can provide a frequency offset higher than half of the electrical bandwidth of the interferogram signal (to avoid frequency folding). This is the case in figure 1 of the main text. However, they can also use separate pump lasers, which is demonstrated in the main text from figure 2 to figure 3. The modulators are driven by amplified electrical signals at frequency close to 22 GHz that are synchronized by electrical phase shifters. The output power of the electrical amplifiers is 33 dBm. The phase modulated pump is then coupled to an intensity modulator to select only portions of the waveform with a uniform chirp. The intensity modulator is driven by the recycled microwave signal from the external termination port of the first phase modulator. The modulation intensity and phase are controlled by an electrical attenuator and phase shifter. A programmable line-by-line waveshaper is used to flatten the EO comb optical spectrum and to nullify the linear chirping so as to form a transform-limited sinc-shaped temporal pulse. The average power from the waveshaper output is around 100 $\mu$W. The EO pulses are amplified by an EDFA before combining with the microresonator signal. 

In the interferogram measurement, the microcavity signal and the EO pulses are combined in a 90/10 coupler and are then detected by a fast photodetector with 50 GHz bandwidth. An FBG filter is used to block the pump laser of the microcavity to avoid saturation in the photodetector. All photodetected signals are recorded using a 4 GHz bandwidth, 20 GSa/s sampling rate oscilloscope.

\section*{Supplementary Note 2: Simulation of soliton formation}

\begin{figure}[!ht]
\includegraphics[width=16.0cm]{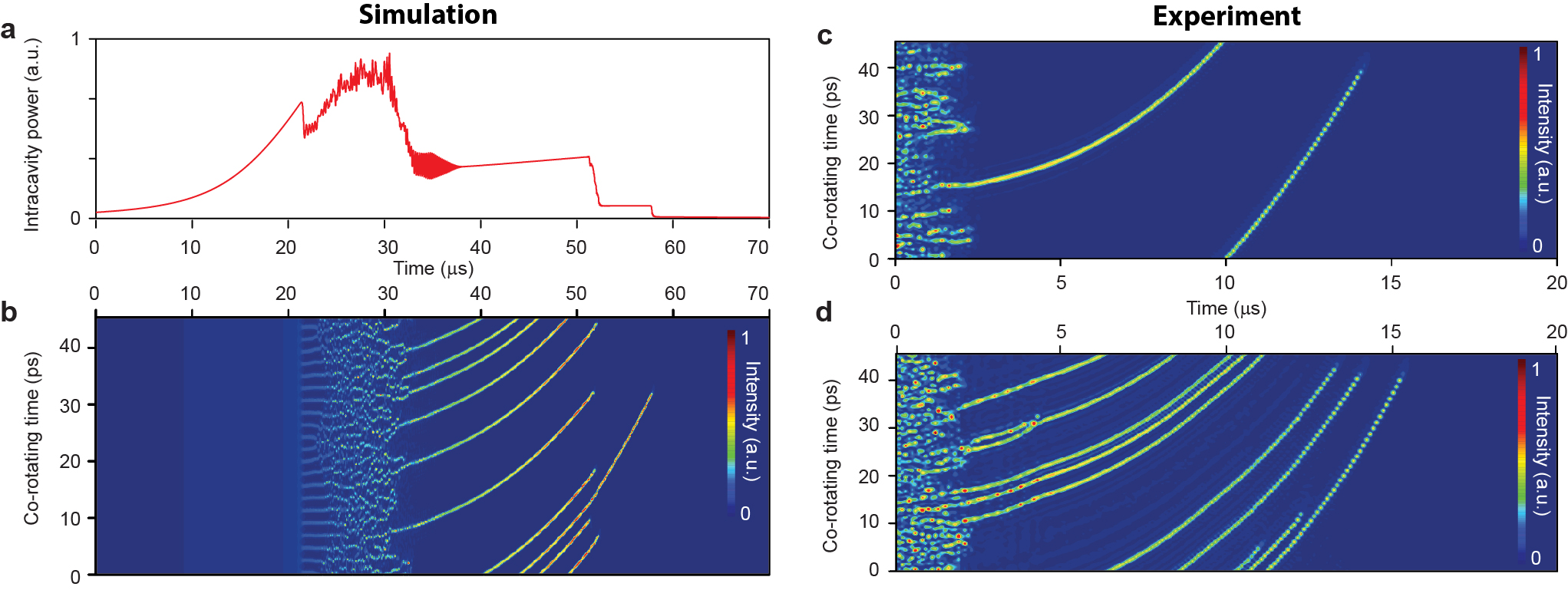}
\caption{{\bf Simulation and measurement of microcavity soliton formation.} {\bf a.} Simulated intracavity power plotted versus time as the pumping laser is tuned across a cavity resonance from higher to lower frequencies. The step features correspond to the formation of solitons. {\bf b.} Simulation results corresponding to panel a and showing the formation of multiple solitons. The slow (horizontal) and co-rotating (vertical) time axes are defined in the main text. In the simulation, the Raman effect and avoided mode crossing are included. {\bf c, d.} Measured soliton trajectories. The frame rate is 10 MHz for these measurements and the resonator is the same one described in the main text. 
}
\label{simulation}
\end{figure}

The soliton formation process is governed by the Lugiato-Lefever equation (LLE) \cite{slugiato1987spatial} augmented by Raman \cite{skarpov2016raman,syi2016theory} and avoided mode crossing\cite{sherr2014mode} effects. The formation process can be simulated numerically using the split-step method \cite{sagrawal2007nonlinear}. The simulated intracavity power and temporal profile are presented in figure S2 (panels a and b, respectively). The time domain result is plotted in the slow and co-rotating time frame. In the simulation, the laser frequency is linearly scanned from higher to lower frequency. For comparison, two measurement results showing soliton formation are presented in figure S2 (c) and S2 (d). Concerning the vertical axis scale, it is noted that because the periodicity of the soliton interferogram signals varies by less than 1 $\%$, the vertical co-rotating time axis can be readily mapped into soliton angular position axis within the circular microcavity as shown in fig. 1d-e in the main text.

\section*{Supplementary Note 3: Measurement of relative soliton position}

The soliton positions can be extracted from the measurement by a peak-finding algorithm. One soliton is selected to be the reference and is always positioned at the zero point of the angular position so as to eliminate the change in soliton repetition rate. The angular position is defined from $- \pi$ to $\pi$. Four representative results are shown in figure S3. In the measurement, the laser frequency is scanned from high to low frequency. In panel (a) and (b), the solitons stabilize relative to each other within a few $\mu$s after formation. In panel (c), the relative soliton positions stabilize immediately after soliton formation. In panel (d), the relative soliton positions stabilize from 9 to 22 $\mu$s and are then observed to destabilize. Note that at some point in time the solitons in all panels are annihilated when the laser tunes beyond the existence detuning range.

\begin{figure}[!ht]
\includegraphics[width=16.0cm]{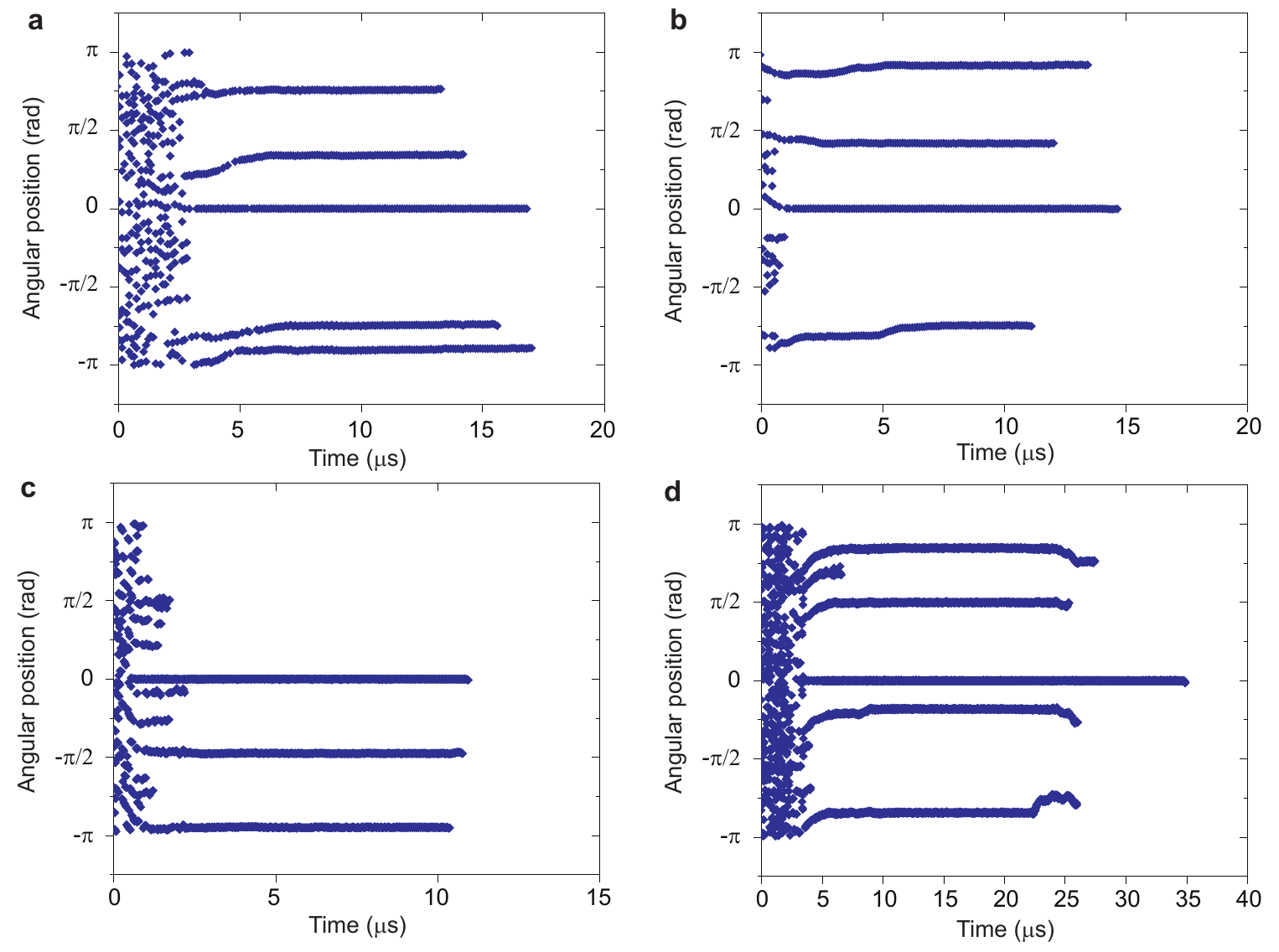}
\caption{{\bf Measurement of relative soliton positions in multiple soliton states.} The positions of each soliton are measured relative to a reference soliton (located at angular position zero) and are plotted versus scan time. The sampling rate for panel {\bf a} and {\bf b} is 10 MHz, while the rate is 50 MHz for panel {\bf c} and {\bf d}. The laser frequency is scanned from high to low frequency for all panels. 
}
\label{position}
\end{figure}

\section*{Supplementary Note 4: Numerical simulation of soliton collision}
Numerical simulation is used to reproduce soliton collisions using the method described in Note 2 above. Four simulation results are shown in figure S4. Note the appearance of the soliton ``splash'' at points of annihilation. This phenomena is noted for observations presented in the main text. 

\begin{figure}[!ht]
\includegraphics[width=16.0cm]{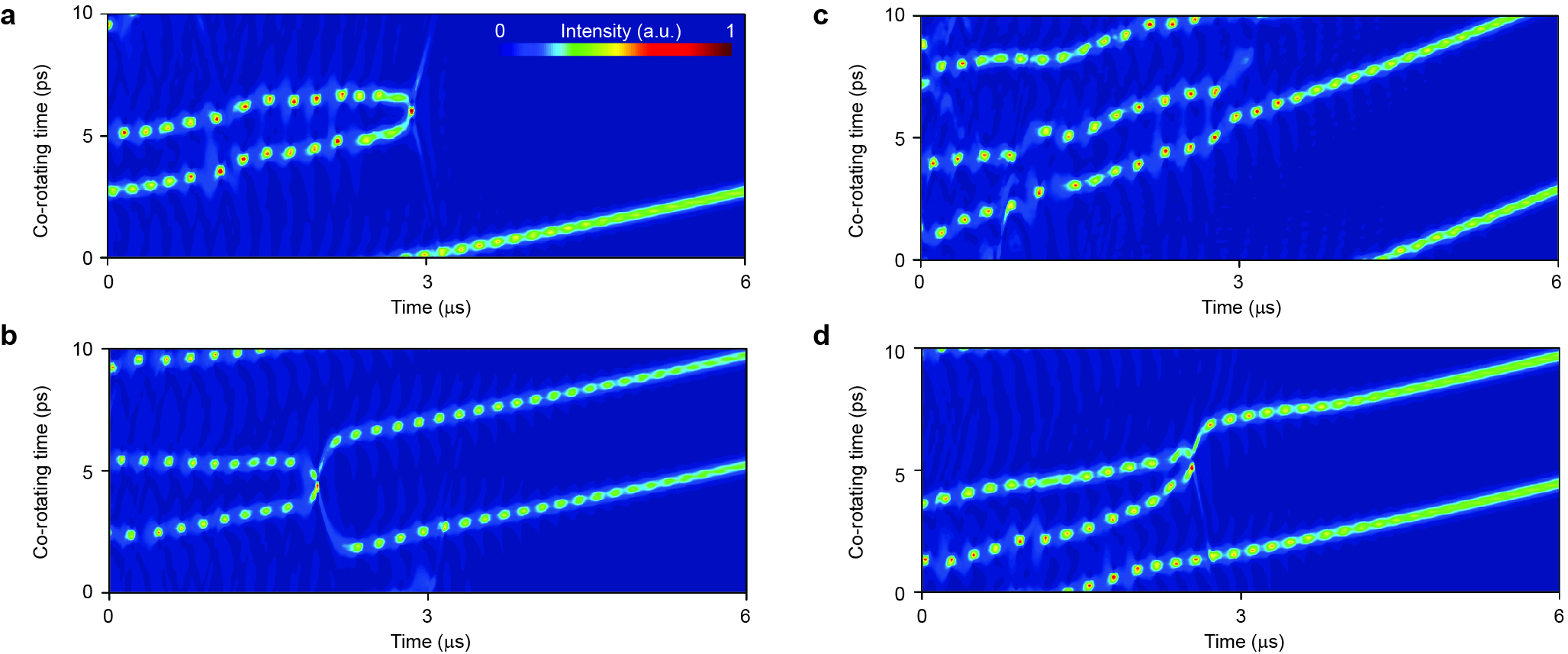}
\caption{{\bf Numerical simulation showing transient soliton scattering events.} {\bf a,} Two solitons collide and annihilate. A soliton ``splash'' appears after annihilation. {\bf b,} Two solitons survive a collision. {\bf c,} Two solitons collide and merge into one soliton. {\bf d,} One soliton hops in location when another soliton is annihilated. Parameters are set similar to experimental condition.}
\label{collision}
\end{figure}

\section*{Supplementary Note 5: Numerical simulation of soliton annihilation}

Numerical simulation (described in Note 2 above) is used to examine soliton properties during annihilation. The laser frequency scans from higher to lower frequency and when the cavity-laser detuning frequency exceeds the soliton existence range, the soliton begins to decay. The calculated soliton amplitude ($A_E$) and pulse width ($\tau_s$) from the simulation are plotted in figure S5 (a). Their product $\tau_s A_E$ is shown to be approximately constant in figure S5 (b). An oscillation of the parameters is seen when the soliton amplitude decays to a small value. 

\begin{figure}[!ht]
\centering
\includegraphics[width=10.0cm]{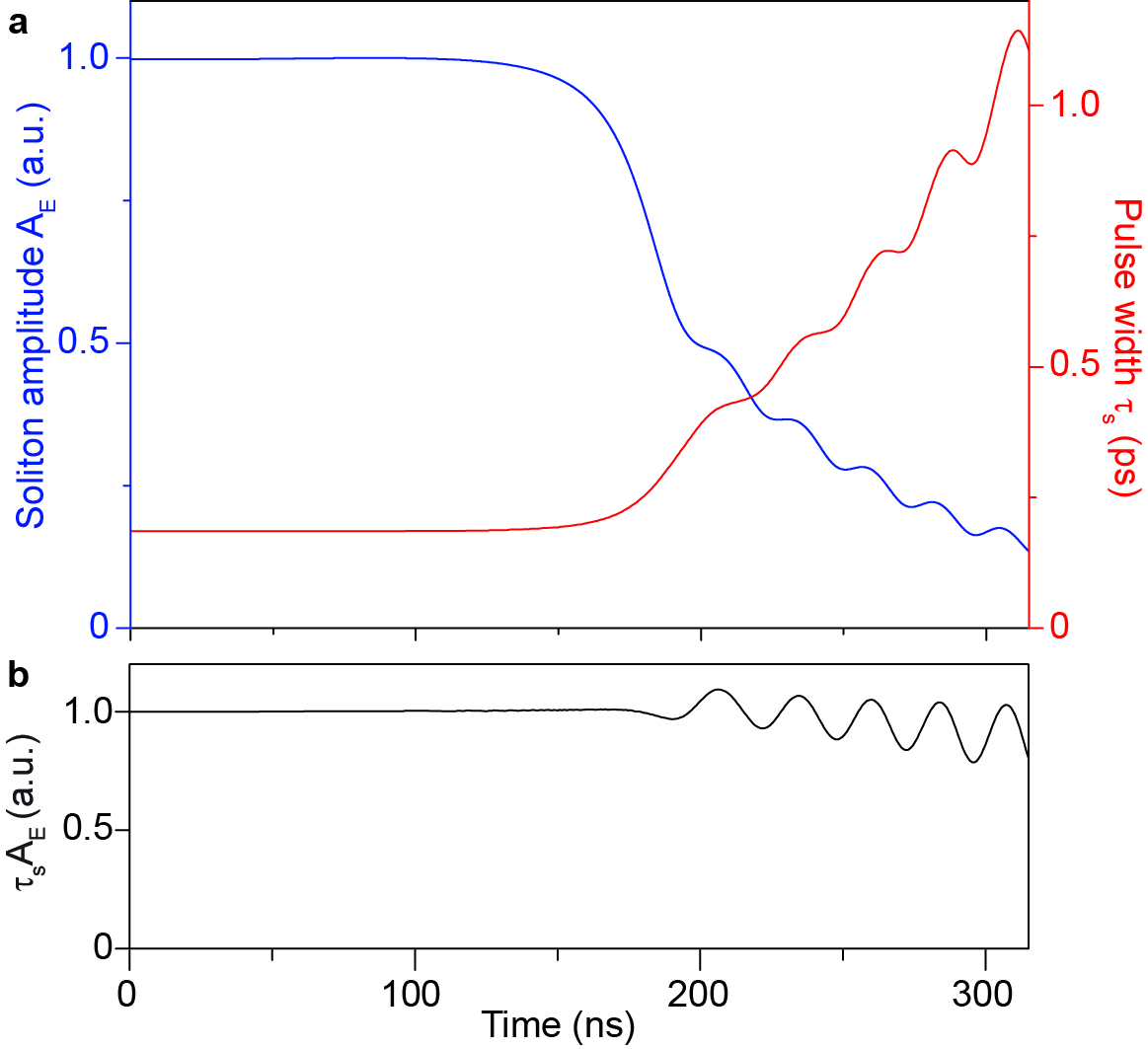}
\caption{{\bf Numerical simulation of soliton annihilation.} The pump laser frequency scans towards longer wavelength and eventually exceeds the soliton existence range. {\bf a.} The simulated soliton amplitude and pulse width during soliton decay. {\bf b.} The product of soliton amplitude and pulse width. The Q-factor in the simulation is set to 100 million.}
\label{annihilation}
\end{figure}

\end{document}